\documentclass{article}
\usepackage{booktabs}
\usepackage{makecell}
\usepackage{subcaption}

\usepackage[utf8]{inputenc}     
\usepackage[english]{babel}     
\usepackage[a4paper, left=1in, right=1in, top=1.25in, bottom=1.25in]{geometry}
\usepackage{graphicx}

\usepackage{amsmath,amssymb}    
\usepackage{amsthm}             
\usepackage{mathtools}          
\usepackage{enumitem}           
\usepackage[hidelinks]{hyperref} 
\usepackage{float}
\usepackage{amscd}
\usepackage{mathrsfs}
\usepackage{setspace}
\usepackage{physics}
\usepackage{xcolor}
\usepackage{tikz}
\usepackage[toc,page,title,titletoc,header]{appendix}
\usepackage{url}
\usepackage{csquotes}
\usepackage[f]{esvect}

\newcommand{\R}{\mathbb{R}}

\newcommand{\E}{\mathbb{E}}
\newcommand{\C}{\mathbb{C}}

\newcommand{\Ha}{\mathcal{H}_{\mathrm{ann}}}
\newcommand{\Hc}{\mathcal{H}_{\mathrm{q}}}
\newcommand{\HaN}{\mathcal{H}_{\mathrm{ann},N}}
\newcommand{\HcN}{\mathcal{H}_{\mathrm{q},N}}
\newcommand{\g}{\mathfrak{g}}
\newcommand{\p}{\mathbb{P}}

\newcommand{\one}{\mathbf{1}}

\let\phi=\varphi
\let\epsilon=\varepsilon

\theoremstyle{plain}
\newtheorem{theorem}{Result}[section]

\newtheorem{conjecture}[theorem]{Conjecture}

\theoremstyle{definition}
\newtheorem{definition}[theorem]{Definition}
\newtheorem{example}[theorem]{Example}

\theoremstyle{remark}
\newtheorem{remark}[theorem]{Remark}

\newenvironment{remarks}{%
  \par\vspace{1ex}%
  \noindent\textbf{Remarks.}\begin{itemize}\setlength\itemsep{2pt}}%
  {\end{itemize}\par\vspace{1ex}}

\usepackage{authblk} 

\title{Phase transitions in non-Hermitian spherical integrals}

\author[1,2]{Pierre Bousseyroux\thanks{Email: pierre.bousseyroux@polytechnique.edu}}
\author[3]{Marc Potters}

\affil[1]{Econophysics Lab, Institut Louis Bachelier, 28 Pl. de la Bourse, Palais Brongniart, 75002 Paris, France}
\affil[2]{LadHyX, UMR CNRS 7646, Ecole Polytechnique, Institut Polytechnique de Paris, 91128 Palaiseau, France}
\affil[3]{Capital Fund Management, Paris, France}

\begin{document}

\maketitle

\begin{abstract}

We study the large-\(N\) asymptotics of a constrained spherical integral for non-Hermitian random matrices, in which the norms and mutual scalar product of two vectors are fixed. In the delocalized regime, the asymptotics are governed by the non-Hermitian transforms \(\mathcal R_1\) and \(\mathcal R_2\) introduced in \cite{BousseyrouxPotters2026RTransforms}. At saddle-point level, the constrained integral exhibits a transition to a localized regime controlled by the largest singular value of a shifted matrix and by the overlap of its associated left and right singular vectors. Motivated by the Hermitian spherical-integral mechanism and by the Coulomb-gas picture, we formulate conjectures for one-eigenvalue large deviations and boundary fluctuations. Throughout, our analysis is carried out in the spirit of mathematical physics.
\end{abstract}

\section{Introduction}
\label{sec:intro}

\subsection{Extreme large deviations in the Hermitian world: two routes}

Large deviations of extreme eigenvalues of random matrices can be studied by
two complementary methods. The first is based on spherical integrals of
Harish-Chandra--Itzykson--Zuber type
\cite{HarishChandra1957,ItzyksonZuber1980}. In rank one, the central object
is
\begin{equation}
    I_N(\theta,X_N)
    =
    \E_{e}
    \left[
        \exp\left(
            N\theta\langle e,X_Ne\rangle
        \right)
    \right],
    \label{eq:rank_one_spherical_integral}
\end{equation}
where $\theta\geq0$ is a scalar parameter, $X_N$ is an $N\times N$
Hermitian matrix, and $e$ is uniformly distributed on the unit sphere.
Guionnet and Ma\"ida \cite{GuionnetMaida2005}, building on earlier
spherical-integral asymptotics of Guionnet and Zeitouni
\cite{GuionnetZeitouni2002}, showed that if the spectrum of
$X_N$ converges to $\mu$ while $\lambda_{\max}(X_N)\to\rho$, the limit
$J(\theta)=\lim N^{-1}\log I_N$ exists and its derivative exhibits a phase
transition,
\begin{equation}
    \partial_\theta J(\mu,\theta,\rho)
    =
    \begin{cases}
        R_\mu(\theta),
        & \theta<\g_\mu(\rho), \\[0.4em]
        \rho-\dfrac1\theta,
        & \theta\geq \g_\mu(\rho),
    \end{cases}
    \label{eq:H_transform_phase_transition}
\end{equation}
where
$\g_\mu(z)=\int_{\R}(z-t)^{-1}\,\mathrm d\mu(t)$ is the Stieltjes transform
and $R_\mu(\theta)=\g_\mu^{-1}(\theta)-1/\theta$ the $R$-transform. For small
$\theta$ the integral only sees the bulk;
above the critical value of $\theta$ the saddle point \emph{sticks to the extreme
eigenvalue}, and the answer explicitly involves $\rho$. This transition is
precisely what makes spherical integrals a useful tool for one-particle
large deviations: it has been used for rank-one and finite-rank deformations
\cite{Maida2007,BenaychGeorgesGuionnetMaida2012}, for sums of matrices in
generic position \cite{GuionnetMaida2020,MergnyPotters2022}, and for Wigner
matrices with sharp sub-Gaussian tails
\cite{GuionnetHusson2020,AugeriGuionnetHusson2021}, i.e.\ in situations
where no eigenvalue Coulomb gas is directly available. This spherical route
is the one we aim to extend to the non-Hermitian setting.

The second route is the classical potential-theoretic, or Coulomb-gas, one.
For invariant Hermitian ensembles
\begin{equation}
    \mathrm d\p_N(M)
    =
    \frac1{Z_N}
    \exp\left\{-N\Tr V(M)\right\}\mathrm dM,
\end{equation}
where $V:\R\to\R$ is the confining potential, $Z_N$ is the normalization
constant, and $\mathrm dM$ denotes Lebesgue measure on the space of
Hermitian matrices. The eigenvalues form a one-dimensional logarithmic gas whose empirical
measure $\mu_N$ converges to the equilibrium measure $\mu_V$, minimizer of
the logarithmic energy \cite{BenArousGuionnet1997,saff2013logarithmic}. In
this picture, pushing the largest eigenvalue to $x$ beyond the right edge
$r_V$ of the support is a \emph{one-particle} event, occurring at speed $N$:
one charge is extracted from the equilibrium sea while the others barely
move \cite{DeanMajumdar2006,MajumdarVergassola2009}. Its cost is the
effective potential
\begin{equation}
    \Psi_V(x)
    =
    V(x)
    -
    2\int_{\R}\log|x-y|\,\mathrm d\mu_V(y)
    -
    \ell_V ,
    \qquad \Psi_V(r_V)=0 ,
    \label{eq:hermitian_coulomb_rate}
\end{equation}
where $\ell_V$ is the normalization constant fixed by $\Psi_V(r_V)=0$.

\subsection{The non-Hermitian problem}

Let $\lambda_1(\mathbf M_N),\ldots,\lambda_N(\mathbf M_N)$ denote the complex
eigenvalues of an $N\times N$ non-Hermitian random matrix $\mathbf M_N$, and
let
\begin{equation}
\mu_N
=
\frac{1}{N}
\sum_{i=1}^N
\delta_{\lambda_i(\mathbf M_N)}
\end{equation}
be its empirical spectral measure, where $\delta_z$ denotes the Dirac mass at
$z\in\C$. We assume that $\mu_N$ converges to a deterministic probability
measure $\mu$ with compact, connected support
$S=\operatorname{supp}\mu\subset\C$. We assume that $S$ is connected in
order to avoid additional geometric complications.

The non-Hermitian analogue of a right large deviation is the event that at
least one eigenvalue lies near a prescribed point $z\notin S$. The cost of
this one-particle excursion is expected to be of order $N$, with rate
$\Psi(z)$:
\begin{equation}
\lim_{\varepsilon\downarrow0}
\lim_{N\to\infty}
-\frac1N\log
\p\left(
\exists i:
\lambda_i(\mathbf M_N)\in B(z,\varepsilon)
\right)
=
\Psi(z),
\label{eq:one_particle_ldp_intro}
\end{equation}
where
$B(z,\varepsilon)=\{w\in\C:|w-z|<\varepsilon\}$ and $\Psi$ is the
one-particle rate function.

When the eigenvalues are distributed according to a two-dimensional
Coulomb-gas law
\begin{equation}
P_N(\lambda_1,\ldots,\lambda_N)
\propto
\exp\left\{
-N\sum_{i=1}^N
V(\lambda_i,\bar\lambda_i)
+
2\sum_{i<j}
\log|\lambda_i-\lambda_j|
\right\},
\label{eq:coulomb_joint_form}
\end{equation}
where $V(z,\bar z)$ is the confining potential of the Coulomb-gas law,
as in the complex Ginibre
\cite{Ginibre1965,BenArousZeitouni1998,rider2003limit}
and elliptic Ginibre
\cite{girko1986elliptic,SommersCrisantiSompolinskyStein1988,
LeeRiser2016,ByunLeeOh2026}
ensembles, the rate function follows from the electrostatic equilibrium
problem.

Define the logarithmic potential of the limiting measure by
\begin{equation}
U_\mu(z)
=
2\int_\C
\log|z-w|\,
\mathrm d\mu(w).
\label{eq:log_potential_intro}
\end{equation}
For $z\notin S$, its Wirtinger derivative is the Stieltjes transform
\begin{equation}
\g_\mu(z)
=
\partial_z U_\mu(z)
=
\int_\C
\frac{\mathrm d\mu(\lambda)}
{z-\lambda},
\label{eq:stieltjes_transform}
\end{equation}
and $\partial_{\bar z}U_\mu(z)=\overline{\g_\mu(z)}$. Here
\begin{equation}
\partial_z
=
\frac{1}{2}
\left(
\partial_{\Re z}
-
i\,\partial_{\Im z}
\right),
\qquad
\Delta=4\,\partial_z\partial_{\bar z}
\label{eq:wirtinger_def}
\end{equation}
is the Wirtinger derivative and the plane Laplacian.

The equilibrium conditions follow from the Euler--Lagrange equations for the
logarithmic energy \cite{saff2013logarithmic}: there exists a constant
$\ell_V$ such that
\begin{equation}
V(z,\bar z)-U_\mu(z)
=
\ell_V
\quad\text{on }S,
\qquad
V(z,\bar z)-U_\mu(z)
\geq
\ell_V
\quad\text{on }\C.
\label{eq:equilibrium_conditions}
\end{equation}
By the Poisson equation of two-dimensional electrostatics, the equilibrium
area density on $S$ is obtained from the bulk potential:
\begin{equation}
\rho_\mu(z)
=
\frac{1}{\pi}\,\partial_z\partial_{\bar z}\,V(z,\bar z)
=
\frac{1}{\pi}\,\partial_z\partial_{\bar z}\,U_\mu(z),
\qquad z\in S,
\label{eq:equilibrium_density_laplacian}
\end{equation}
in the interior of the support, at points where these derivatives are
defined. Since $V=U_\mu+\ell_V$ on $S$, equivalently
$\Delta U_\mu=4\pi\rho_\mu$ on the support.

If one eigenvalue is constrained near $z\notin S$, while the remaining
eigenvalues retain the equilibrium distribution $\mu$ at leading order, its
interaction with the bulk contributes
\begin{equation}
N V(z)
-
2\sum_{j=2}^N
\log|z-\lambda_j|
=
N\left[
V(z,\bar z)-U_\mu(z)
\right]
+o(N).
\end{equation}
Subtracting the equilibrium value $\ell_V$ gives the one-particle rate
function
\begin{equation}
\Psi_V(z)
=
V(z,\bar z)-U_\mu(z)-\ell_V,
\qquad z\notin S.
\label{eq:one_particle_potential_rate}
\end{equation}
This formula compares the effective energy at $z$ with its value on the
equilibrium support.

For the complex Ginibre ensemble,
\begin{equation}
V(z,\bar z)=|z|^2,
\qquad
\mathrm d\mu(z)
=
\frac{1}{\pi}
\mathbf 1_{\{|z|\leq1\}},
\mathrm d^2z,
\end{equation}
so that $S$ is the unit disk. By rotational symmetry, the logarithmic
potential depends only on $r=|z|$ and is given by
\begin{equation}
U_\mu(r)
=
\begin{cases}
r^2-1,
& 0\leq r\leq1, \\[2mm]
2\log r,
& r\geq1.
\end{cases}
\label{eq:ginibre_logarithmic_potential}
\end{equation}
Consequently,
\begin{equation}
V(r)-U_\mu(r)=1,
\qquad 0\leq r\leq1,
\end{equation}
and hence $\ell_V=1$. Outside the unit disk,
\begin{equation}
\Psi_{\mathrm{Gin}}(r)
=
r^2-1-2\log r,
\qquad r\geq1.
\label{eq:ginibre_rate_intro}
\end{equation}
This computation is standard; see, for example,
\cite{Ginibre1965,BenArousZeitouni1998,ByunForrester2024}.

In this paper, $\mathbf M_N$ is called \emph{rotationally invariant} if
\begin{equation}
\mathbf U\mathbf M_N\mathbf U^*
\overset{\mathrm d}{=}
\mathbf M_N
\qquad
\text{for every deterministic unitary matrix }\mathbf U,
\label{eq:rotational_invariance_intro}
\end{equation}
and \emph{bi-invariant} if
\begin{equation}
\mathbf U\mathbf M_N\mathbf V
\overset{\mathrm d}{=}
\mathbf M_N
\qquad
\text{for every pair of deterministic unitary matrices }
\mathbf U,\mathbf V,
\label{eq:bi_invariance_intro}
\end{equation}
where $\overset{\mathrm d}{=}$ denotes equality in distribution. Neither
invariance property alone implies that the eigenvalues have a Coulomb-gas
density of the form \eqref{eq:coulomb_joint_form}. A particular class of
bi-invariant ensembles is instead defined by a matrix density
\begin{equation}
\mathrm d\p_N(\mathbf M)
\propto
\exp\left\{
-N\Tr W(\mathbf M\mathbf M^*)
\right\}
\mathrm d\mathbf M.
\label{eq:bi_potential_model_intro}
\end{equation}
Here $\mathrm d\mathbf M$ denotes Lebesgue measure on the space of complex
$N\times N$ matrices.
In such models, $W$ governs the singular values of $\mathbf M$, but it should
not in general be identified with an eigenvalue potential $V$.

The Coulomb-gas derivation of \eqref{eq:one_particle_potential_rate} requires
an explicit eigenvalue density of the form \eqref{eq:coulomb_joint_form}. For
rotationally or bi-invariant non-Hermitian models, and more generally whenever
the eigenvalue law is not directly available, the Hermitian lesson of
\eqref{eq:H_transform_phase_transition} suggests a complementary route: study
instead the asymptotics of non-Hermitian spherical integrals and read off large
deviations from the phase transition of their saddle point. This paper
develops this complementary route by deriving the non-Hermitian transition
at saddle-point level and formulating conjectural consequences for
one-eigenvalue large deviations.

\subsection{Main message and organization}

We study the constrained non-Hermitian spherical integral of
Section~\ref{sec:setting}. At saddle-point level it exhibits a transition
analogous to \eqref{eq:H_transform_phase_transition}: below the transition
the asymptotics are governed by $\mathcal R_1$ and $\mathcal R_2$; at the
transition the regular saddle reaches the largest singular value of a
shifted matrix, and beyond it the integration vectors concentrate on the
corresponding left and right singular vectors
(Result~\ref{thm:main_transition}).

Section~\ref{sec:conjectures} formulates conjectures for one-eigenvalue
large deviations and boundary fluctuations, motivated by the Hermitian
mechanism and by the Coulomb-gas picture. The paper is organized as
follows. Section~\ref{sec:setting} sets notations and recalls the objects
of \cite{BousseyrouxPotters2026RTransforms}. Section~\ref{sec:main} states
the saddle-point transition and the conjectures.
Section~\ref{sec:examples} works out the null, Ginibre and elliptic cases.
The saddle-point derivations are gathered in the appendices.

\section{Setting and notations}
\label{sec:setting}

Throughout, we adopt a physicist's approach to large-$N$ asymptotics
and saddle-point arguments. $\mathbf M=\mathbf M_N$ is an $N\times N$
complex matrix, deterministic or random, $\one=\one_N$ is the identity
matrix, and
$\langle x,y\rangle=\sum_{i=1}^N\bar x_i y_i$ is the standard scalar product
on $\C^N$. For a sequence of matrices $\mathbf C_N$, we use
$\tau[f(\mathbf C)]:=\lim_{N\to\infty}N^{-1}\Tr f(\mathbf C_N)$ whenever
the limit exists.

For every $z\in\C$, set
\begin{equation}
    \mathbf B_z := \mathbf M-z\one ,
    \qquad
    s_1(z)\geq s_2(z)\geq\dots\geq s_N(z)
\end{equation}
where the singular values are the non-negative square roots of the
eigenvalues of $\mathbf B_z\mathbf B_z^*$. We write
$s_{\max,N}(z)=s_1(z)$. Whenever the limit exists,
\begin{equation}
    \sigma_{\max}(z)
    :=
    \lim_{N\to\infty}s_{\max,N}(z)
    \label{eq:sigma_max_def}
\end{equation}
denotes the limiting largest singular value of $\mathbf B_z$, whether it
belongs to the bulk edge or to an isolated singular branch.

\subsection{Non-Hermitian spherical integrals}

We first recall the annealed $\mathcal H$-transform introduced in
\cite{BousseyrouxPotters2026RTransforms}. We write it $\Ha$ in what
follows.

\begin{definition}[Annealed $\mathcal H$-transform \cite{BousseyrouxPotters2026RTransforms}]
\label{def:H}
Let $\mathbf M$ be a deterministic or random $N\times N$ matrix, and let
$\psi_1,\psi_2\in\C^N$. Let $\mathbf U$ be an independent Haar-distributed
unitary matrix. We define
\begin{equation}
\frac{1}{2N}\,
\log
\E_{\mathbf M,\mathbf U}\!\left[
\exp\!\Bigl(2N\,\Re\langle \psi_1,\, \mathbf{U}\mathbf{M}\mathbf{U}^* \psi_2\rangle\Bigr)
\right]
\;=\;
\HaN\bigl(
\|\psi_1\|\,\|\psi_2\|,\,
\langle\psi_1,\psi_2\rangle
\bigr).
\end{equation}
We set $\Ha:=\lim_{N\to\infty}\HaN$ whenever this limit exists.
The reference vectors may be chosen so that
$\|\psi_1\|^2=\|\psi_2\|^2=\alpha$ and
$\langle\psi_1,\psi_2\rangle=\beta$.
The associated transforms are
\begin{equation}\label{eq:R_transforms_def}
\mathcal{R}_1(\alpha,\beta) := \partial_{\alpha}\,\Ha(\alpha,\beta),
\qquad
\mathcal{R}_2(\alpha,\beta) := 2\,\partial_{\beta}\,\Ha(\alpha,\beta),
\end{equation}
where $\partial_\beta=\tfrac12(\partial_{\Re\beta}-i\,\partial_{\Im\beta})$
is the Wirtinger derivative with respect to $\beta$.
\end{definition}

If $\mathbf M$ is rotationally invariant, the Haar average is redundant
and $\E_{\mathbf M,\mathbf U}$ may equivalently be replaced by
$\E_{\mathbf M}$. For deterministic $\mathbf M$, the expectation over
$\mathbf M$ is of course trivial.

\begin{remarks}
\item By the Cauchy--Schwarz inequality, only the region $|\beta|\leq\alpha$
is accessible; it will be convenient to use the ratio
$x=\beta/\alpha$, which satisfies $|x|\leq1$.
\item We write $x=\beta/\alpha\in\C$, with $|x|<1$. The saddle-point
equations are covariant under simultaneous rotations of $\mathbf M$ and
$\beta$ whenever the ensemble has this symmetry.
\end{remarks}

We now introduce its quenched constrained counterpart, defined for a
fixed realization of the matrix. We write it $\Hc$.

\begin{definition}[Quenched constrained spherical integral]
\label{def:Hc}
For $\alpha>0$ and $\beta\in\C$ with $|\beta|<\alpha$, let
\begin{equation}
    e^{2N\,\HcN(\alpha,\beta)}
    :=
    \frac{1}{Z_N(\alpha,\beta)}
    \int_{\C^N\times\C^N}
    e^{2N\Re\langle \phi_1,\mathbf M\phi_2\rangle}
    \,
    \delta\bigl(\|\phi_1\|^2-\alpha\bigr)
    \,
    \delta\bigl(\|\phi_2\|^2-\alpha\bigr)
    \,
    \delta_{\C}\bigl(\langle\phi_1,\phi_2\rangle-\beta\bigr)
    \,\mathrm d\phi_1\,\mathrm d\phi_2 ,
\end{equation}
where $Z_N(\alpha,\beta)$ is the same integral without the exponential
factor, and $\Hc:=\lim_{N\to\infty}\HcN$. Here $\delta$ is the Dirac
distribution on $\R$, $\delta_\C(w)=\delta(\Re w)\delta(\Im w)$, and
$\mathrm d\phi=\prod_{j=1}^N\mathrm d(\Re\phi_j)\mathrm d(\Im\phi_j)$ is
Lebesgue measure on $\C^N$.
\end{definition}

The two objects are related by
\begin{equation}
e^{2N\,\HaN(\alpha,\beta)}
=
\E_{\mathbf M}
\left[
e^{2N\,\HcN(\alpha,\beta)}
\right],
\label{eq:annealed_quenched_relation}
\end{equation}
where the dependence of $\HcN$ on the realization of $\mathbf M$ is
implicit.

\subsection{Singular-value resolvents and overlaps}
\label{subsec:resolvents}

The asymptotics of $\Hc$ are expressed through functions associated with
the singular values of $\mathbf B_z$. At finite $N$, for
$\omega>s_{\max,N}(z)$, define
\begin{align}
\g_{1,N}(\omega, z)
&:=
\frac1N\Tr\!\left[\omega\bigl(\omega^2 \one-\mathbf B_z\mathbf B_z^*\bigr)^{-1}\right],
\label{eq:g1_def}
\\[0.3em]
\g_{2,N}(\omega, z)
&:=
\frac1N\Tr\!\left[\bigl(\omega^2 \one-\mathbf B_z\mathbf B_z^*\bigr)^{-1}\mathbf B_z^*\right],
\label{eq:g2_def}
\end{align}
together with
\begin{equation}
    \mathcal L_N(\omega,z)
    := \frac{1}{N}\log \det\bigl(\omega^2\one - \mathbf B_z\mathbf B_z^*\bigr) .
    \label{eq:L_def}
\end{equation}
Whenever their large-$N$ limits exist, they are denoted by
$\g_1$, $\g_2$ and $\mathcal L$, respectively.
Note that $\partial_\omega\mathcal L = 2\g_1$ and
$\partial_{\bar z}\mathcal L = \overline{\g_2}$.

To see what these functions measure, introduce the singular value
decomposition
\begin{equation}
    \mathbf B_z
    =
    \sum_{i=1}^N s_i(z)\,
    \ket{u_i(z)}\bra{v_i(z)} ,
\end{equation}
with left and right singular vectors $\ket{u_i}$, $\ket{v_i}$, and define the
\emph{left--right singular overlaps}
\begin{equation}
    p_i(z) := \braket{u_i(z)}{v_i(z)} ,
    \qquad |p_i(z)|\leq 1 .
    \label{eq:overlap_def}
\end{equation}
Then, at finite $N$,
\begin{equation}
    \g_{1,N}(\omega,z)
    =
    \frac1N\sum_{i=1}^N
    \frac{\omega}{\omega^2-s_i(z)^2},
    \qquad
    \g_{2,N}(\omega,z)
    =
    \frac1N\sum_{i=1}^N
    \frac{s_i(z)\,p_i(z)}{\omega^2-s_i(z)^2} .
    \label{eq:g1g2_svd}
\end{equation}
Thus $\g_{1,N}$ only depends on the singular values of $\mathbf B_z$, while
$\g_{2,N}$ also knows about the relative geometry of the left and right singular
bases. Three elementary consequences will matter.

\begin{enumerate}[label=(\alph*),itemsep=2pt]
\item \emph{Domination.} For $\omega>s_{\max,N}(z)$, the coefficients of
$\g_{1,N}$ are positive, $s_i\leq\omega$ and $|p_i|\leq1$, whence
\begin{equation}
    |\g_{2,N}(\omega,z)|\;\leq\;\g_{1,N}(\omega,z),
    \qquad \omega>s_{\max,N}(z) .
    \label{eq:g2_le_g1}
\end{equation}
\item \emph{Hermitian degeneration.} If $\mathbf M$ is Hermitian and $z=0$,
then $\mathbf B_0=\mathbf M$, the singular vectors coincide with the
eigenvectors up to signs, and
$p_i=\operatorname{sgn}(\lambda_i)\in\{\pm1\}$: the overlaps carry no
additional information. For genuinely non-Hermitian matrices they are
nontrivial, and this is precisely the new difficulty of the non-Hermitian
theory.
\item \emph{Behaviour near the largest singular value.} If the largest singular value
$s_{\max,N}(z)$ is simple, then as $\omega\downarrow s_{\max,N}(z)$ both sums in
\eqref{eq:g1g2_svd} are dominated by the top term, so that
\begin{equation}
    \frac{\g_{2,N}(\omega,z)}{\g_{1,N}(\omega,z)}
    \;\xrightarrow[\omega\downarrow s_{\max,N}(z)]{}\;
    p_1(z) .
    \label{eq:ratio_overlap}
\end{equation}
The ratio of the two resolvents near their first pole therefore gives the
overlap of the corresponding singular vectors. This observation will be
used to locate the transition.
\end{enumerate}

\section{Main results}
\label{sec:main}

This section states the phase transition of the constrained spherical
integral---the non-Hermitian analogue of \eqref{eq:H_transform_phase_transition}---and
records conjectures on one-eigenvalue large deviations suggested by the
Hermitian spherical-integral route.

\subsection{The phase transition of the constrained integral}

At saddle-point level, the constrained integral has a two-phase
structure. We call the saddle \emph{delocalized} when it lies in the
interior of the admissible domain, $\omega>\sigma_{\max}(z)$, and
\emph{localized} when it lies on its boundary,
$\omega=\sigma_{\max}(z)$. The transition occurs when the interior saddle
reaches this boundary.

We assume that the admissible interior saddle is unique whenever it
exists.

\begin{theorem}[Saddle-point transition]
\label{thm:main_transition}
Take $\beta=x\alpha$ with $x\in\C$ and $|x|<1$. Define
\begin{equation}
    \mathcal F(\omega,z;\alpha,\beta)
    :=
    \omega\alpha
    +\Re( z\beta)
    -\tfrac12\,\mathcal L(\omega,z)
    -\tfrac12\log\bigl(\alpha^2-|\beta|^2\bigr)
    -1 .
    \label{eq:master_functional}
\end{equation}
Then $\Hc(\alpha,\beta)=\mathcal F(\omega^*,z^*;\alpha,\beta)$,
where the pair $(\omega^*,z^*)$ is determined as follows.
\begin{enumerate}[label=\emph{(\Roman*)},itemsep=2pt]
\item \emph{Delocalized phase.} If the system
\begin{equation}
    \g_1(\omega,z)=\alpha,
    \qquad
    \g_2(\omega,z)=\beta
    \label{eq:saddle_system}
\end{equation}
admits a solution with $\omega>\sigma_{\max}(z)$, then
$(\omega^*,z^*)$ is the unique admissible solution. In this phase
\begin{equation}
    \partial_\alpha\Hc=\mathcal R_1(\alpha,\beta),
    \qquad
    2\,\partial_\beta\Hc=\mathcal R_2(\alpha,\beta) .
\end{equation}
\item \emph{Localized phase.} Otherwise, there is no admissible interior
solution and $(\omega^*,z^*)$ is the minimizer of
$\mathcal F(\,\cdot\,;\alpha,\beta)$ on the boundary
$\{\omega=\sigma_{\max}(z)\}$.
\end{enumerate}
In both phases,
\begin{equation}
    \partial_\alpha\Hc
    =
    \omega^*-\frac{1}{(1-|x|^2)\,\alpha},
    \qquad
    2\,\partial_\beta\Hc
    =
    z^*+\frac{\bar x}{(1-|x|^2)\,\alpha} .
    \label{eq:derivative_formulas}
\end{equation}
In the localized phase, the large-$\alpha$ saddle is selected by the
largest singular value of a shifted matrix:
\begin{equation}
    \zeta_x
    \in
    \operatorname*{argmin}_{z\in\C}
    \bigl[\sigma_{\max}(z)+\Re(zx)\bigr],
    \qquad
    p(\zeta_x)=x,
    \qquad
    \sigma_x:=\sigma_{\max}(\zeta_x).
    \label{eq:overlap_condition}
\end{equation}
In particular
\begin{equation}
    \partial_\alpha\Hc
    \;\xrightarrow[\alpha\to\infty]{}\;
    \sigma_x,
    \qquad
    2\,\partial_\beta\Hc
    \;\xrightarrow[\alpha\to\infty]{}\;
    \zeta_x .
\end{equation}
\end{theorem}

\begin{remarks}
\item The distinction between $\Ha$ and $\Hc$ is important. The former is
annealed, whereas the latter is the quenched spherical integral for a
typical realization. In the delocalized phase connected to the origin,
\begin{equation}
    \partial_\alpha\Hc=\mathcal R_1=\partial_\alpha\Ha,
    \qquad
    2\,\partial_\beta\Hc=\mathcal R_2=2\,\partial_\beta\Ha.
\end{equation}
Since $\Hc(0,0)=\Ha(0,0)=0$, one obtains
\begin{equation}
    \Hc(\alpha,\beta)=\Ha(\alpha,\beta).
\end{equation}
throughout this component: $\Ha$ and $\Hc$ coincide below the
transition. Beyond the transition, $\Hc$
follows the localized saddle and may differ from $\Ha$.
\item For large $\alpha$, the exponential weight favors pairs of unit
vectors satisfying the overlap constraint and maximizing
$\Re\langle u,\mathbf Mv\rangle$. Thus
\begin{equation}
    \Hc(\alpha,x\alpha)
    \;\sim\;
    \alpha\,
    \max\Bigl\{
    \Re\langle u,\mathbf Mv\rangle
    \;:\;
    \|u\|=\|v\|=1,\;
    \langle u,v\rangle=x
    \Bigr\},
\end{equation}
and introducing a complex multiplier $\zeta$ for
$\langle u,v\rangle=x$ leads directly to the shifted matrix
$\mathbf B_\zeta=\mathbf M-\zeta\one$. The maximizing vectors are its top
left and right singular vectors.
\item Differentiating the top singular-value curve gives
$p(z)=-2\,\partial_z\sigma_{\max}(z)$, so the stationarity condition is
$p(\zeta_x)=x$. First-order perturbation theory yields
$\delta s_{\max,N}=-\Re\{\delta z\langle u_1,v_1\rangle\}$, hence
\begin{equation}
    p(z):=\lim_{N\to\infty}\braket{u_1(z)}{v_1(z)}
    =
    -2\,\partial_z \sigma_{\max}(z),
    \label{eq:overlap_limit}
\end{equation}
where $\partial_z=\tfrac12(\partial_{\Re z}-i\partial_{\Im z})$ as in
\eqref{eq:wirtinger_def}. For a bi-invariant law, $\sigma_{\max}(z)$
depends only on $|z|$, and
$p(z)=-e^{-i\arg z}\,\sigma_{\max}'(|z|)$.
\label{rem:overlap_from_s}
\end{remarks}

\subsection{Conjectures}
\label{sec:conjectures}

Motivated by the Hermitian spherical-integral mechanism, the Coulomb-gas
picture, and the examples below, we formulate two conjectures for
one-eigenvalue large deviations and maximal boundary excursions.

Our constrained spherical-integral approach suggests a possible way of
probing a single complex eigenvalue excursion outside the bulk. One may hope
that, along a suitable path in parameter space, the regular saddle reaches the
boundary and undergoes a sticking transition, giving rise to a localized branch
encoding such an excursion. Starting from the finite-$N$ integral
representations, one may then formally interchange the matrix expectation with
the auxiliary integrations and investigate the corresponding large-$N$
continuation. This possible route is currently under investigation.

The conjectures are intended for a generic regime. A natural class that we are
currently investigating as a testing ground is provided by non-Hermitian
ensembles
\begin{equation}
    \mathrm d\p_N(\mathbf M)
    \propto
    \exp\left\{
        -N\operatorname{Tr}W(\mathbf M,\mathbf M^*)
    \right\}
    \mathrm d\mathbf M,
    \label{eq:nonhermitian_test_class}
\end{equation}
where $W$ is a self-adjoint noncommutative polynomial and the integral
defining $\mathrm d\p_N$ is assumed to converge. The two conjectures below
are stated for a generic ensemble of this form.

\subsubsection{Large deviations of one complex eigenvalue}
\label{subsec:ld}

\begin{conjecture}[One-eigenvalue large deviations]
\label{thm:ld}
Let $\mathbf M_N$ be an ensemble of the form
\eqref{eq:nonhermitian_test_class}. Suppose that its
limiting spectral support $S$ is connected and that the bulk branch of the
electrostatic potential, viewed as a function of the independent variables
$z$ and $z^*=\bar z$, admits an analytic continuation
$V_{\mathrm{bulk}}$ to the exterior domain under consideration, defined by
\begin{equation}
    V_{\mathrm{bulk}}(z,\bar z)=U_\mu(z),
    \qquad z\in S,
    \label{eq:bulk_equilibrium_relation}
\end{equation}
with $U_\mu$ as in \eqref{eq:log_potential_intro}. Here $V_{\mathrm{bulk}}$
is defined by this continuation; it is not a confining potential from a
Coulomb-gas density. Then, wherever this continuation is defined, for
$z\notin S$,
\begin{equation}
    \Psi(z)
    =
    V_{\mathrm{bulk}}(z,\bar z)-U_\mu(z).
    \label{eq:ld_theorem_formula}
\end{equation}
Thus $\Psi$ is the difference between the analytic continuation of the
bulk branch and the electrostatic potential on its true exterior branch.
\end{conjecture}

\begin{remark}[Coulomb-gas and Hermitian consistency]
For ensembles whose eigenvalues form a two-dimensional Coulomb gas with
confining potential $Q$, the equilibrium relation
\begin{equation}
    Q-U_\mu=\ell_Q
    \qquad\text{on }S
\end{equation}
implies that the continuation of the bulk branch is
$V_{\mathrm{bulk}}=Q-\ell_Q$. Hence Conjecture~\ref{thm:ld} reduces to the
standard one-particle effective potential
\begin{equation}
    \Psi(z)=Q(z,\bar z)-U_\mu(z)-\ell_Q.
\end{equation}
The same argument on the real line recovers the usual one-particle rate
function for invariant Hermitian ensembles. The conjectural content of
\eqref{eq:ld_theorem_formula} is that this effective-potential construction
continues to hold for non-Hermitian invariant ensembles without
assuming an explicit Coulomb-gas eigenvalue density.
\end{remark}

\begin{remark}[Conjectural consequence: radial spectral measure]
\label{cor:radial}
If $\mu$ is radial, with support $\{R_-\leq|z|\leq R_+\}$, then
$\g_\mu(w)=1/w$ for $|w|>R_+$ and $\g_\mu(w)=0$ for $|w|<R_-$. Then
\begin{align}
    \Psi_{\mathrm{out}}(r)
    &=
    V_{\mathrm{bulk}}(r)-V_{\mathrm{bulk}}(R_+)-2\log\frac{r}{R_+},
    \qquad r\geq R_+ ,
    \label{eq:radial_outer}\\
    \Psi_{\mathrm{in}}(r)
    &=
    V_{\mathrm{bulk}}(r)-V_{\mathrm{bulk}}(R_-),
    \qquad\qquad\qquad\;\; r\leq R_- .
    \label{eq:radial_inner}
\end{align}
\end{remark}

For Ginibre, the radial consequence of Conjecture~\ref{thm:ld} recovers
\eqref{eq:ginibre_rate_intro}.

\subsubsection{Maximal distance from the spectral support}
\label{sec:maximal_distance}

\begin{conjecture}[Maximal distance from the support]
\label{thm:maximal_distance}
Let $\mathbf M_N$ be an ensemble of the form
\eqref{eq:nonhermitian_test_class}. Suppose that its connected limiting support
$S\subset\C$ has a regular boundary, and that the limiting spectral density
$\rho_\mu$ is continuous and stays
positive on $\partial S$. Write
$\lambda_1,\ldots,\lambda_N$ for its
eigenvalues and set
\begin{equation}
    D_N
    =
    \max_{1\leq j\leq N}
    \operatorname{dist}(\lambda_j,S).
    \label{eq:maximal_distance_definition}
\end{equation}
Then
\begin{equation}
    D_N
    \underset{N\to+\infty}{\sim}
    \sqrt{
        \frac{\log N}
        {4\pi N\rho_*}
    },
    \qquad
    \rho_*
    :=
    \min_{z\in\partial S}\rho_\mu(z).
    \label{eq:maximal_distance_final}
\end{equation}
\end{conjecture}

The heuristic argument below explains the scaling of
Conjecture~\ref{thm:maximal_distance}. It is motivated by
Conjecture~\ref{thm:ld}. If a single eigenvalue is placed at $z\notin S$ while
the others remain in equilibrium, its atypical probability is of order
$\exp\{-N\Psi(z)\}$ at speed~$N$. The regularity assumptions imply that
$V_{\mathrm{bulk}}-U_\mu$ and its first derivatives match the constant
zero function
at $\partial S$. Outside $S$, $U_\mu$ is harmonic, whereas
\eqref{eq:equilibrium_density_laplacian} gives
$\Delta V_{\mathrm{bulk}}=4\pi\rho_\mu$ at the boundary. Since $\Psi$ vanishes identically along
the boundary, its tangential second derivative also vanishes there.
Consequently its quadratic expansion is entirely expressed through the
distance $d:=\operatorname{dist}(z,S)$:
\begin{equation}
    \Psi(z)
    =
    2\pi\,\rho_\mu(z_0)\,d^2
    +
    o(d^2),
    \qquad
    z_0\in\partial S,\;
    \operatorname{dist}(z,S)=d,
    \label{eq:psi_distance_expansion}
\end{equation}
where $\rho_\mu$ is the limiting spectral density. Inserting this into the
large-deviation weight gives
\begin{equation}
    \exp\left\{
        -N\Psi(z)
    \right\}
    =
    \exp\left\{
        -2\pi N\,\rho_\mu(z_0)\,d^2+o(Nd^2)
    \right\}.
    \label{eq:local_distance_tail}
\end{equation}
For Ginibre, \eqref{eq:ginibre_rate_intro} is consistent with
\eqref{eq:psi_distance_expansion} since $\rho_\mu=1/\pi$ on the unit disk. We use
only this leading exponential behavior in what follows.

\paragraph{Number of candidate eigenvalues.}

The $N$ eigenvalues occupy a domain of order-one area, so their typical
microscopic spacing is of order $N^{-1/2}$. An eigenvalue lying deep inside
$S$ cannot cross the boundary through a fluctuation on this scale. The
relevant candidates are therefore the eigenvalues contained in a boundary
layer of width $N^{-1/2}$.

If $L=|\partial S|$ denotes the boundary length, this layer has area of order
$L N^{-1/2}$. Since the macroscopic eigenvalue density is of order $N$, the
number of eigenvalues in the layer is of order
\begin{equation}
    N\times L N^{-1/2}
    \asymp
    \sqrt{N}.
    \label{eq:number_boundary_candidates}
\end{equation}
Thus, at the level of this heuristic, approximately $\sqrt{N}$ eigenvalues can
produce the extreme excursion. They are not independent, but the argument
assumes that correlations along the boundary do not change the leading
logarithmic balance.

\paragraph{Uniform boundary density.}

Assume first that
\begin{equation}
    \rho_\mu(z)=\rho_0,
    \qquad z\in\partial S.
\end{equation}
For each candidate, a distance greater than $d$ is exponentially equivalent
to $\exp\{-2\pi N\rho_0 d^2\}$. The expected number of such excursions is
therefore estimated as
\begin{equation}
    \sqrt{N}\,
    \exp\left\{
        -2\pi N\rho_0 d^2
    \right\}.
\end{equation}
The maximal distance is located at the threshold where this quantity becomes
of order one:
\begin{equation}
    \sqrt{N}\,
    \exp\left\{
        -2\pi N\rho_0 D_N^2
    \right\}
    \asymp
    1.
    \label{eq:extreme_balance}
\end{equation}
Taking logarithms gives
\begin{equation}
    \tfrac12\log N
    -
    2\pi N\rho_0 D_N^2
    \asymp
    0,
\end{equation}
and hence, as $N\to\infty$,
\begin{equation}
    D_N
    \sim
    \sqrt{
        \frac{\log N}
        {4\pi N\rho_0}
    }.
    \label{eq:maximal_distance_constant_mu}
\end{equation}
The factor $\tfrac12$ multiplying $\log N$ reflects the fact that the effective
number of candidates is of order $\sqrt{N}$ rather than $N$.

\paragraph{Non-uniform boundary density.}

Suppose now that $\rho_\mu$ varies along the boundary, and set
$\rho_*:=\min_{z\in\partial S}\rho_\mu(z)$.
An excursion of a given size is exponentially favored near the set
\begin{equation}
    \mathcal M
    =
    \left\{
        z\in\partial S:
        \rho_\mu(z)=\rho_*
    \right\}.
\end{equation}
At first logarithmic order, the previous balance therefore gives, as
$N\to\infty$,
\begin{equation}
    D_N
    \sim
    \sqrt{
        \frac{\log N}
        {4\pi N\rho_*}
    }.
    \label{eq:maximal_distance_variable_mu}
\end{equation}

\section{Examples}
\label{sec:examples}

\subsection{The null matrix: normalization and complex-overlap check}

\begin{example}[$\mathbf M=0$]
\label{ex:null}
This example fixes the normalization and checks the complex-overlap
dependence. For $\mathbf M=0$ one has $\Hc\equiv0$ identically. Since
$\mathbf B_z\mathbf B_z^*=|z|^2\one$,
\begin{equation}
    \g_1=\frac{\omega}{\omega^2-|z|^2},
    \qquad
    \g_2=\frac{-\bar z}{\omega^2-|z|^2},
    \qquad
    \mathcal L=\log(\omega^2-|z|^2),
\end{equation}
and the system \eqref{eq:saddle_system} with
$x=\beta/\alpha\in\C$ gives
\begin{equation}
    h=\frac{1}{\alpha(1-|x|^2)},
    \qquad
    \omega^*=h,
    \qquad
    z^*=-\bar x\,h.
\end{equation}
Thus $\g_2(\omega^*,z^*)=\alpha x=\beta$. Substituting in
\eqref{eq:master_functional}, the $\mathcal L$-term cancels the entropy term
exactly and $\mathcal F=(1-|x|^2)/(1-|x|^2)-1=0$. Both derivative formulas
\eqref{eq:derivative_formulas} vanish identically. There is no localized
phase: the singular spectrum of $\mathbf B_z$ is a point mass and the saddle
never sticks.
\end{example}

\subsection{Complex Ginibre: explicit saddle transition}
\label{subsec:ginibre_example}

This subsection works out the constrained-saddle transition. Let
$\mathbf M$ be complex Ginibre, with $\E|M_{ij}|^2=1/N$. The annealed
$\mathcal H$-transform was computed in
\cite{BousseyrouxPotters2026RTransforms}:
\begin{equation}
    \Ha(\alpha,\beta)=\frac{\alpha^2}{2},
    \qquad
    \mathcal R_1(\alpha,\beta)=\alpha,
    \qquad
    \mathcal R_2(\alpha,\beta)=0.
\end{equation}
We write $x=r e^{i\varphi}$ with $r=|x|<1$. The modulus $r$ controls the
transition; the phase $\varphi$ fixes the direction of the shifted saddle.

\paragraph{Singular data of the shifted matrix.}
The singular-value functions of $\mathbf B_z=\mathbf M-z\one$ are, with
$h=\omega-\g_1$ and $t=|z|^2$,
\begin{equation}
    \g_1=\frac{h}{h^2-t},
    \qquad
    \g_2=\frac{-\bar z}{h^2-t},
    \qquad
    \omega=h+\frac{h}{h^2-t}.
    \label{eq:ginibre_selfconsistent}
\end{equation}
By bi-invariance, $\sigma_{\max}(z)$ depends only on $q=|z|$. The upper
singular edge is reached when $\mathrm d\omega/\mathrm dh=0$, i.e.\ on
$(h^2-t)^2=h^2+t$. The positive root is
\begin{equation}
    w(q)=\sqrt{1+8q^2},
    \qquad
    h_e(q)^2=\frac{2q^2+1+w(q)}{2},
\end{equation}
and substituting gives
\begin{equation}
    \Sigma(q)
    :=
    \sigma_{\max}(z)
    =
    h_e(q)\,\frac{3+w(q)}{1+w(q)},
    \qquad
    P(q)
    :=
    \frac{2q\,h_e(q)}{h_e(q)^2+q^2}.
    \label{eq:ginibre_overlap_formula}
\end{equation}
For $z\ne0$,
$p(z)=-e^{-i\arg z}\,P(|z|)$, while $p(0)=0$. Along
$z=-q e^{-i\varphi}$,
\begin{equation}
    \mathbf B_z=\mathbf M+q e^{-i\varphi}\one,
    \qquad
    p(z)=e^{i\varphi}P(q).
    \label{eq:ginibre_ray_overlap}
\end{equation}
At $t=0$ one recovers $h_e=1$ and $\sigma_{\max}(0)=2$.

\begin{figure}[t]
\centering
\includegraphics[width=\textwidth]{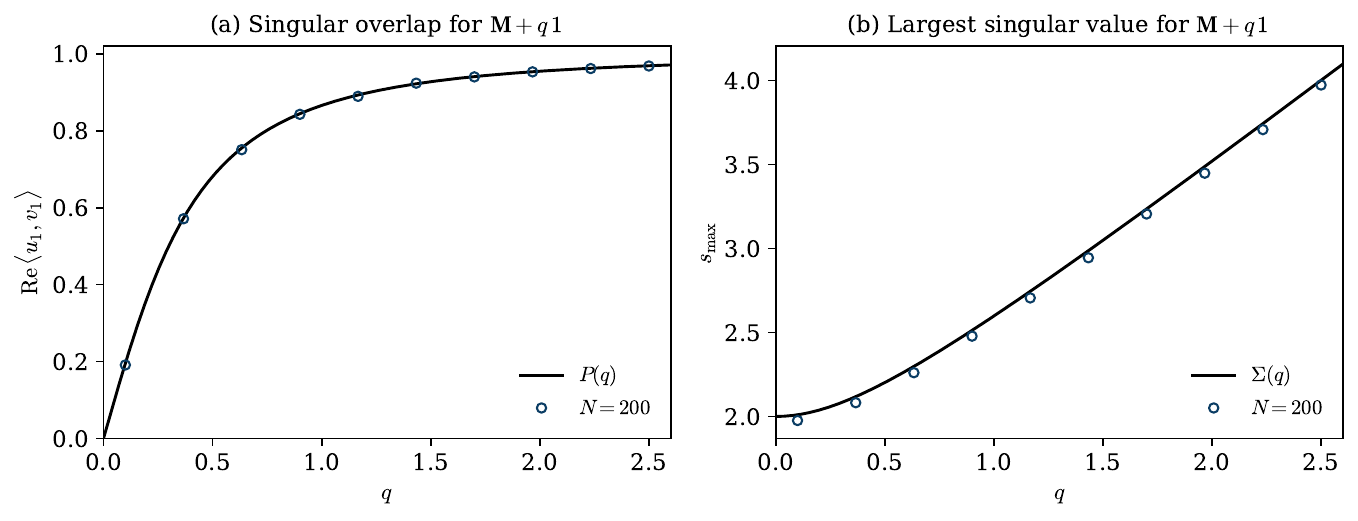}
\caption{Singular-edge data for Ginibre, $\mathbf M+q\,\one$, at $N=200$.
(a) Mean $\Re\langle u_1,v_1\rangle$ compared with $P(q)$.
(b) Mean top singular value compared with $\Sigma(q)$.
Averages over $12$ samples.}
\label{fig:overlap}
\end{figure}

\paragraph{Explicit $\Hc$.}
Define
\begin{equation}
    \alpha_c(r)=\frac{1}{\sqrt{1+r^2}}.
    \label{eq:ginibre_alpha_c}
\end{equation}
Then
\begin{equation}
    \Hc(\alpha,x\alpha)
    =
    \begin{cases}
        \dfrac{\alpha^2}{2},
        & \alpha\leq\alpha_c(r), \\[0.8em]
        \displaystyle
        \min_{q\geq0}\,
        \mathcal F\!\bigl(\Sigma(q),\,-q e^{-i\varphi};\,\alpha,\,x\alpha\bigr),
        & \alpha>\alpha_c(r),
    \end{cases}
    \label{eq:ginibre_Hc_piecewise}
\end{equation}
where $\mathcal F$ is defined in \eqref{eq:master_functional}.

\emph{Delocalized branch.}
For $\alpha\leq\alpha_c(r)$ the saddle of \eqref{eq:saddle_system} is
\begin{equation}
    h=\frac{1}{\alpha(1-r^2)},
    \qquad
    \omega^*=\alpha+h,
    \qquad
    z^*=-\bar x\,h .
    \label{eq:ginibre_saddle}
\end{equation}
Substituting in \eqref{eq:master_functional} gives
$\Hc=\Ha=\alpha^2/2$. One has
$\partial_\alpha\Hc=\mathcal R_1=\alpha$ and
$2\partial_\beta\Hc=\mathcal R_2=0$, so $\Ha$ and $\Hc$ coincide below
the transition. For $\alpha>\alpha_c(r)$, $\Hc$ follows the localized
saddle while $\Ha=\alpha^2/2$ remains the annealed value.

\emph{Transition.}
The delocalized saddle reaches the singular edge when
$\omega^*=\sigma_{\max}(z^*)$. Inserting $t=r^2h^2$ in
$(h^2-t)^2=h^2+t$ yields $h_c^2=(1+r^2)/(1-r^2)^2$, and therefore
\begin{equation}
    \omega_c=\frac{2}{(1-r^2)\sqrt{1+r^2}},
    \qquad
    z_c=-\,\bar x\,\frac{\sqrt{1+r^2}}{1-r^2}.
    \label{eq:ginibre_transition}
\end{equation}
At $x=0$ one recovers $\alpha_c=1$, $\omega_c=2$, $z_c=0$. Increasing
$r$ lowers $\alpha_c(r)$.

\emph{Localized branch.}
For $\alpha>\alpha_c(r)$, the saddle sticks to
$\{\omega=\Sigma(q),\,z=-q e^{-i\varphi}\}$. Since
$\mathcal L(\Sigma(q),z)$ is radial, write
$\Lambda(q):=\mathcal L(\Sigma(q),z)$ for $|z|=q$. Then
\begin{equation}
    \Hc(\alpha,x\alpha)
    =
    \min_{q\geq0}
    \Bigl\{
    \alpha\bigl[\Sigma(q)-qr\bigr]
    -\tfrac12\,\Lambda(q)
    -\tfrac12\log\bigl(\alpha^2(1-r^2)\bigr)
    -1
    \Bigr\},
    \label{eq:ginibre_Hc_localized}
\end{equation}
The minimizer $q^*(\alpha,r)$ moves along the edge from the touching point
$q_c=|z_c|$ at $\alpha=\alpha_c(r)$ towards the localization radius $q_r$
defined by
\begin{equation}
    P(q_r)=r,
    \qquad
    \zeta_x=-q_r e^{-i\varphi},
    \qquad
    \sigma_x=\Sigma(q_r).
    \label{eq:ginibre_overlap_condition}
\end{equation}
By \eqref{eq:ginibre_ray_overlap}, this is precisely
$p(\zeta_x)=x$. The radial overlap at the touching point satisfies
$P(q_c)=2r/(1+r^2)>r$ for $0<r<1$, so the saddle slides before reaching
$q_r$. In the large-$\alpha$ limit,
\begin{equation}
    \partial_\alpha\Hc
    \;\xrightarrow[\alpha\to\infty]{}\;
    \sigma_x,
    \qquad
    2\,\partial_\beta\Hc
    \;\xrightarrow[\alpha\to\infty]{}\;
    \zeta_x=-q_r e^{-i\varphi}.
\end{equation}
As $r\to1$ at fixed $\varphi$, the constraint forces
$v=e^{i\varphi}u$, and the optimization reduces to the Hermitian matrix
\begin{equation}
\frac{e^{i\varphi}\mathbf M+e^{-i\varphi}\mathbf M^*}{2}.
\end{equation}
Indeed,
\begin{equation}
\Re\langle u,\mathbf Mv\rangle
=
\left\langle
u,\frac{e^{i\varphi}\mathbf M+e^{-i\varphi}\mathbf M^*}{2}u
\right\rangle,
\end{equation}
while $q_r\to\infty$.

\emph{Check at $r=0$.}
For zero prescribed overlap, symmetry keeps the minimizer at $q=0$ and
\eqref{eq:ginibre_Hc_piecewise} gives
\begin{equation}
    \Hc(\alpha,0)
    =
    \begin{cases}
        \alpha^2/2, & \alpha\leq1,\\
        2\alpha-\log\alpha-\tfrac32, & \alpha>1 .
    \end{cases}
    \label{eq:ginibre_Hc_x0}
\end{equation}
The second branch is obtained by evaluating
$\mathcal F(2,0;\alpha,0)$ at the upper Ginibre singular edge
$\sigma_{\max}(0)=2$.

\subsection{Elliptic Ginibre: complex continuation and one-particle rate}
\label{subsec:elliptic_example}

For the complex elliptic Ginibre ensemble with parameter
$\tau\in[0,1)$, the entries are centered complex Gaussian variables with
covariance
\begin{equation}
\E[M_{ij}\overline{M_{kl}}]
=
\frac1N\delta_{ik}\delta_{jl},
\qquad
\E[M_{ij}M_{kl}]
=
\frac{\tau}{N}\delta_{il}\delta_{jk}.
\end{equation}
This example tests Conjecture~\ref{thm:ld} in the complex plane, rather
than solving the constrained-saddle transition. The annealed transform is
$\Ha(\alpha,\beta)=\tfrac12\alpha^2+\tfrac\tau2\Re(\beta^2)$, whence
\begin{equation}
    \mathcal R_1(\alpha,\beta)=\alpha,
    \qquad
    \mathcal R_2(\alpha,\beta)=\tau\beta .
\end{equation}
Its limiting support is
\begin{equation}
    S_\tau
    =
    \left\{
        z\in\C:
        \frac{(\Re z)^2}{(1+\tau)^2}
        +
        \frac{(\Im z)^2}{(1-\tau)^2}
        \leq
        1
    \right\},
    \label{eq:elliptic_support}
\end{equation}
with uniform density
\begin{equation}
    \rho_\tau=\frac{1}{\pi(1-\tau^2)}.
\end{equation}
For $z\in\C\setminus S_\tau$, the exterior Stieltjes transform is
\begin{equation}
    \g_\mu(z)=\theta(z),
    \qquad
    \theta(z)
    =
    \frac{z-\sqrt{z^2-4\tau}}{2\tau},
    \label{eq:elliptic_stieltjes}
\end{equation}
where the square-root branch is chosen so that
$\sqrt{z^2-4\tau}\sim z$ at infinity. Equivalently,
$z=\theta^{-1}+\tau\theta$. At $\tau=0$, this is understood as
$\theta(z)=1/z$.

To make the analytic continuation explicit, first regard $z$ and
$w$ as independent complex variables and set
\begin{equation}
    \widehat V_{\mathrm{bulk}}(z,w)
    =
    \frac{zw-\frac{\tau}{2}(z^2+w^2)}{1-\tau^2}-1 .
    \label{eq:elliptic_bulk_continuation}
\end{equation}
On the physical slice $w=\bar z$, this gives
\begin{equation}
    V_{\mathrm{bulk}}(z,\bar z)
    =
    \frac{|z|^2-\tau\Re(z^2)}{1-\tau^2}-1 .
\end{equation}
The corresponding exterior continuation is
\begin{equation}
    \widehat U_{\mathrm{ext}}(z,w)
    =
    -\log\bigl(\theta(z)\theta(w)\bigr)
    +\frac{\tau}{2}\bigl(\theta(z)^2+\theta(w)^2\bigr).
\end{equation}
On the physical slice $w=\bar z$, the exterior branch is
\begin{equation}
    U_\mu(z)
    =
    \widehat U_{\mathrm{ext}}(z,\bar z)
    =
    -\log|\theta(z)|^2+\tau\Re\bigl(\theta(z)^2\bigr),
    \qquad z\notin S_\tau .
    \label{eq:elliptic_exterior_potential}
\end{equation}
Conjecture~\ref{thm:ld} therefore gives the two-dimensional rate
\begin{equation}
    \Psi_\tau(z)
    =
    \frac{|z|^2-\tau\Re(z^2)}{1-\tau^2}
    -1
    +\log|\theta(z)|^2
    -\tau\Re\bigl(\theta(z)^2\bigr),
    \qquad z\notin S_\tau .
    \label{eq:elliptic_rate}
\end{equation}

\emph{Real-axis check.}
For $a>1+\tau$, \eqref{eq:elliptic_rate} becomes
\begin{equation}
    \Psi_\tau(a)
    =
    \frac{a^2}{1+\tau}
    -1
    +2\log\theta(a)
    -\tau\theta(a)^2,
    \qquad
    \theta(a)=\frac{a-\sqrt{a^2-4\tau}}{2\tau}.
    \label{eq:elliptic_rate_real_axis}
\end{equation}
This agrees with the rightmost-eigenvalue rate function of
\cite{ByunLeeOh2026}. The limit $\tau\to0$ gives the Ginibre expression.

\paragraph{Boundary excursions.}
For fixed $\tau<1$, the quadratic expansion of
\eqref{eq:elliptic_rate} at $\partial S_\tau$ gives
\begin{equation}
    \exp\left\{
        -\frac{2N}{1-\tau^2}\,
        \bigl[\operatorname{dist}(z,S_\tau)\bigr]^2
    \right\},
    \label{eq:elliptic_distance_tail}
\end{equation}
in agreement with $2\pi\rho_\tau=2/(1-\tau^2)$. Since $\rho_\tau$ is
constant along the ellipse, Conjecture~\ref{thm:maximal_distance} predicts
\begin{equation}
    D_N
    \sim
    \frac12
    \sqrt{
        \frac{(1-\tau^2)\log N}{N}
    },
    \qquad
    0\leq\tau<1
    \text{ fixed}.
    \label{eq:elliptic_maximal_distance}
\end{equation}
For $\tau=0$, this gives
\begin{equation}
    D_N
    \sim
    \frac12\sqrt{\frac{\log N}{N}},
    \label{eq:ginibre_maximal_distance}
\end{equation}
in agreement with the leading-order behavior of the Ginibre spectral radius
\cite{rider2003limit}. 

\paragraph{Acknowledgments.} This research was conducted within the
Econophysics \& Complex Systems Research Chair, under the aegis of the
Fondation du Risque, the Fondation de l'\'Ecole polytechnique, the
\'Ecole polytechnique and Capital Fund Management.

\bibliographystyle{plain}
\bibliography{References}

\newpage
\begin{appendices}

\section{The delocalized branch: saddle-point derivation}
\label{app:proof_deloc}

We derive \eqref{eq:master_functional} and the saddle
system \eqref{eq:saddle_system}.

\subsection{Constraint representation}

The constraints of Definition~\ref{def:Hc} fix
\begin{equation}
C=\|\phi_1\|^2=\alpha,
\qquad
D=\|\phi_2\|^2=\alpha,
\qquad
E=\langle \phi_1,\phi_2\rangle=\beta .
\end{equation}
We use the contour representation of the Dirac delta,
$\delta(x)=\int_{-i\infty}^{i\infty}\frac{\mathrm dy}{2\pi i}e^{-xy}$, its
complex version being
$\delta_\C(\zeta)=\delta(\Re \zeta)\,\delta(\Im \zeta)$. Denoting by
$I_N(\alpha,\beta)$ the numerator integral of Definition~\ref{def:Hc} and
inserting the three constraints with conjugate multipliers
$(c,d,e)$ rescaled by $N$, we obtain, up to multiplicative constants
independent of $(\alpha,\beta)$,
\begin{align}
I_N(\alpha,\beta)
&\propto
\int
\exp\Bigl\{
N\bigl[
c\|\phi_1\|^2
+
d\|\phi_2\|^2
+
2\Re\langle\phi_1,(\mathbf M+e\one)\phi_2\rangle
\bigr]
-
N\bigl[
c\alpha+d\alpha+2\Re(e \beta)
\bigr]
\Bigr\}
\\
&\qquad{}\times
\mathrm d\phi_1\,\mathrm d\phi_2\,
\mathrm dc\,\mathrm dd\,\mathrm d^2e .
\end{align}

\subsection{Gaussian integration}

Set $\Phi=(\phi_1,\phi_2)^{\mathsf T}\in\C^{2N}$. The quadratic form in the
exponent is $-N\langle\Phi,\mathcal A\Phi\rangle$ with
\begin{equation}
\mathcal A(c,d,e)
=
-\begin{pmatrix}
c\,\one & \mathbf M+e\one\\
(\mathbf M+e\one)^* & d\,\one
\end{pmatrix} .
\label{eq:A_matrix}
\end{equation}
The complex Gaussian identity
$\int_{\C^{2N}}e^{-N\langle\Phi,\mathcal A\Phi\rangle}\mathrm d\Phi
=\pi^{2N}N^{-2N}/\det\mathcal A$ (valid when the Hermitian part of
$\mathcal A$ is positive definite) yields
\begin{equation}
I_N(\alpha,\beta)
\propto
\int
\exp\Bigl\{
-N\,\mathcal S_N
\Bigr\}
\mathrm dc\,\mathrm dd\,\mathrm d^2e ,
\end{equation}
with the action
\begin{equation}
\mathcal S_N
=
c\alpha+d\alpha+2\Re(e \beta)
+
\frac1N\log\det\mathcal A(c,d,e) .
\label{eq:action}
\end{equation}

\subsection{Saddle-point equations}

Write $\mathcal A^{-1}$ in $N\times N$ blocks $G_{ab}$ and let
$\g_{ab}=\frac1N\Tr G_{ab}$. Using
$\partial_t\log\det\mathcal A=\Tr(\mathcal A^{-1}\partial_t\mathcal A)$
one finds
\begin{equation}
\frac1N\partial_c\log\det\mathcal A=-\g_{11},
\quad
\frac1N\partial_d\log\det\mathcal A=-\g_{22},
\quad
\frac1N\partial_e\log\det\mathcal A=-\g_{21},
\quad
\frac1N\partial_{\bar e}\log\det\mathcal A=-\g_{12} .
\end{equation}
The stationarity of \eqref{eq:action} in $(c,d,e,\bar e)$ therefore
imposes
\begin{equation}
\alpha=\g_{11},
\qquad
\alpha=\g_{22},
\qquad
\beta=\g_{21},
\qquad
\bar\beta=\g_{12} ,
\label{eq:saddle_CDE}
\end{equation}
and hence
\begin{equation}
\g_{11}\Big|_{(c_\star,d_\star,e_\star)}=\alpha,
\qquad
\g_{22}\Big|_{(c_\star,d_\star,e_\star)}=\alpha,
\qquad
\g_{21}\Big|_{(c_\star,d_\star,e_\star)}=\beta .
\label{eq:saddle_finite_N}
\end{equation}
with the conjugate equation
$\g_{12}=\bar\beta$. Then
\begin{equation}
\HcN(\alpha,\beta)
=
-\tfrac12\,\mathcal S_N^{\star}
-\tfrac12\log\bigl(\alpha^2-|\beta|^2\bigr)
+\text{const}
+o(1),
\label{eq:Hc_from_action}
\end{equation}

For completeness, the entropy in the normalization can be read from the
$2\times2$ Gram matrix of the two vectors,
\begin{equation}
    \mathsf G=
    \begin{pmatrix}
      \alpha&\beta\\ \bar\beta&\alpha
    \end{pmatrix},
    \qquad
    \det\mathsf G=\alpha^2-|\beta|^2 .
\end{equation}
The volume of pairs of vectors with fixed Gram matrix is proportional, at
exponential order, to $(\det\mathsf G)^N$. Consequently
\begin{equation}
    \frac1N\log Z_N(\alpha,\beta)
    =
    \log\bigl(\alpha^2-|\beta|^2\bigr)
    +\text{constant independent of }(\alpha,\beta),
\end{equation}
which gives the entropy term in \eqref{eq:Hc_from_action}.

\subsection{Dictionary \texorpdfstring{$(c,d,e)\leftrightarrow(\omega,z)$}{(c,d,e) and (omega,z)}}

Since both norm constraints equal $\alpha$, the saddle is symmetric and
we take $c_\star=d_\star=:-\omega$,
and we set $e_\star=:-z$. Then $\mathbf M+e\one=\mathbf B_z$ and
\eqref{eq:A_matrix} becomes
\begin{equation}
\mathcal A
=
\begin{pmatrix}
\omega\,\one & -\mathbf B_z\\
-\mathbf B_z^* & \omega\,\one
\end{pmatrix},
\end{equation}
which is positive definite precisely when
$\omega>s_{\max,N}(z)$: this is the finite-$N$ admissible region of the
multipliers. Its determinant is $\det(\omega^2\one-\mathbf B_z\mathbf B_z^*)$.
Block inversion gives
$G_{11}=\omega(\omega^2-\mathbf B_z\mathbf B_z^*)^{-1}$ and
$G_{21}=\mathbf B_z^*(\omega^2-\mathbf B_z\mathbf B_z^*)^{-1}$, so
that $\g_{11}=\g_{22}=\g_{1,N}(\omega,z)$ and
$\g_{21}=\g_{2,N}(\omega,z)$, while
$\frac1N\log\det\mathcal A=\mathcal L_N(\omega,z)$. After taking the
large-$N$ limit, the saddle equations
\eqref{eq:saddle_finite_N} become \eqref{eq:saddle_system}, including
$\g_2=\beta$ for complex $\beta$, and
\eqref{eq:Hc_from_action} becomes \eqref{eq:master_functional}. The
additive constant is fixed by
Example~\ref{ex:null}. One checks directly that
$\mathcal F$ is stationary in $(\omega,z)$ on the system
\eqref{eq:saddle_system}, using
$\partial_\omega\mathcal L=2\g_1$ and
$\partial_{\bar z}\mathcal L=\overline{\g_2}$, so that the envelope theorem
yields the exact derivative formulas \eqref{eq:derivative_formulas}.

\subsection{Connection with \texorpdfstring{$\mathcal R_1$ and $\mathcal R_2$}{R1 and R2}}

In the delocalized phase, the quenched saddle equations reproduce the
derivatives of the annealed $\mathcal H$-transform. Hence $\Ha$ and
$\Hc$ coincide on the component connected to the
origin. This is the small-constraint annealed--quenched equivalence of
\cite{BousseyrouxPotters2026RTransforms}. The regular solution remains
admissible until it reaches $\omega=\sigma_{\max}(z)$, the large-$N$
limit of the finite-$N$ admissible boundary.

\section{The localized branch}
\label{app:proof_frozen}

When $(\alpha,\beta)$ leaves the delocalized region, the regular stationary
point of $\mathcal F$ in $\{\omega>\sigma_{\max}(z)\}$ no longer exists and
the saddle is constrained by $\omega=\sigma_{\max}(z)$. At finite $N$ the
analogous mechanism is the familiar sticking of a saddle point to the
largest pole of a resolvent, as in the Hermitian spherical integral
\cite{GuionnetMaida2005} and in the finite-$N$ discussion of
\cite[Ch.~10 and 14]{potters2020first}. As a function of $\omega$, the
integrand has singularities at the singular values of $\mathbf B_z$;
starting from the admissible region $\omega>s_1(z)$, the first singularity
encountered is $\omega=s_1(z)$. When the formal saddle reaches this
singularity, the steepest-descent contour can no longer be deformed through
the regular stationary point. At the level of the present saddle-point
calculation it then remains attached to the pole. Two consequences follow.

\subsection{Pole dominance and the overlap}

In the strongly localized regime $\alpha\to\infty$, the saddle equation
$\g_{1,N}=\alpha$ forces the top-pole contribution to dominate the regular
part of \eqref{eq:g1g2_svd}. Hence
\begin{equation}
    \frac{\g_{2,N}(\omega,z)}{\g_{1,N}(\omega,z)}
    \longrightarrow
    p_1(z),
    \qquad
    \frac{\beta}{\alpha}=x,
\end{equation}
and therefore $p(\zeta_x)=x$.

\subsection{Value of the localized saddle}

\emph{(i) Value and derivatives.} The pinned contour contributes the same
functional form $\mathcal F(\omega^*,z^*)$, now evaluated at the
constrained minimizer on $\{\omega=\sigma_{\max}(z)\}$; since the manifold
does not depend on $(\alpha,\beta)$, the envelope theorem still yields
\eqref{eq:derivative_formulas}.

\emph{(ii) Localization and the overlap condition.} As $\alpha\to\infty$ at
fixed $x$, $\mathcal F\sim\alpha\bigl[\omega+\Re(z x)\bigr]$ on the
manifold, so the sticking point converges to a minimizer of
$\sigma_{\max}(z)+\Re(zx)$. By \eqref{eq:overlap_limit},
$p(z)=-2\,\partial_z\sigma_{\max}(z)$, and the stationarity condition is
$p(\zeta_x)=x$. Equivalently, undoing the Legendre transform,
\begin{equation}
    \lim_{\alpha\to\infty}\frac{\Hc(\alpha,x\alpha)}{\alpha}
    =
    \max_{\substack{\|u\|=\|v\|=1\\ \langle u,v\rangle=x}}
    \Re\langle u,\mathbf M v\rangle
    =
    \min_{\zeta\in\C}
    \bigl[
    \sigma_{\max}(\zeta)+\Re(\zeta x)
    \bigr] ,
\end{equation}
where $\sigma_{\max}(\zeta)$ is the limiting top singular value of
$\mathbf B_\zeta=\mathbf M-\zeta\one$. This shows directly why the top
singular pair of the optimally shifted matrix appears. At the microscopic
level, the integration vectors concentrate on that pair: this is the
localization referred to in the main text.

\end{appendices}

\end{document}